\documentclass[iop,numerical,prb,amsmath,amssymb,reprint]{revtex4-1}
\usepackage{graphicx}
\usepackage{dcolumn}
\usepackage{bm}
\usepackage{multirow}
\usepackage{rotating}
\usepackage{booktabs,tabularx,threeparttable}
\usepackage{float}
\usepackage{hyperref}
\usepackage[usenames,dvipsnames,svgnames,table]{xcolor}
\usepackage[normalem]{ulem}

\begin{document}

\title{Light propagation in quasiperiodic dieletric multilayers separated by graphene}
\author{Carlos H. Costa}
\affiliation{Universidade Federal do Cear\'a, Campus Avan\c{c}ado de Russas, Russas-CE 62900-000, Brazil}
\author{Luiz F. C. Pereira}
\affiliation{Departamento de F\'{i}sica Te\'orica e Experimental, Universidade Federal do Rio Grande do Norte, Natal-RN 59078-900, Brazil}
\author{Claudionor G. Bezerra}
\email[Author to whom correspondence should be addressed. Electronic mail:]{cbezerra@fisica.ufrn.br}
\affiliation{Departamento de F\'{i}sica Te\'orica e Experimental, Universidade Federal do Rio Grande do Norte, Natal-RN 59078-900, Brazil}
\date{\today}

\begin{abstract}
The study of photonic crystals, artificial materials whose dielectric properties can be tailored according to the stacking of its constituents, remains an attractive research area.
In this article we have employed a transfer matrix treatment to study the propagation of light waves in Fibonacci quasiperiodic dieletric multilayers with graphene embedded. We calculated their dispersion and transmission spectra in order to investigate the effects of the graphene monolayers and quasiperiodic disorder on the system physical behavior. The quasiperiodic dieletric multilayer is composed of two building blocks, silicon dioxide (building block $A=\textrm{SiO}_{2}$) and titanium dioxide (building block $B=\textrm{TiO}_{2}$). Our numerical results show that the presence of graphene monolayers reduces the transmissivity on the whole range of frequency and induces a transmission gap in the low frequency region. Regarding the polarization of the light wave, we found that the transmission coefficient is higher for the transverse magnetic (TM) case than for the transverse electric (TE) one. We also conclude from our numerical results that the \textit{graphene induced photonic bandgaps} (GIPBGs) do not depend on the polarization (TE or TM) of the light wave nor on the  Fibonacci generation index $n$. Moreover, the GIPBGs are \textit{omnidirectional photonic band gaps}, therefore light cannot propagate in this structures for frequencies lower than a certain value, whatever the incidence angle. Finally, a plot of the transmission spectra versus  chemical potential shows that one can, in principle, adjust the width of the photonic band gap by tuning the chemical potential via a gate voltage.
\end{abstract}

\maketitle 

\section{Introduction}\label{sec:introduction}

The study of photonic crystals (PCs), which are artificial materials whose dielectric properties are subject to design and control according to the stacking pattern of its constituents, started in the 90's, with the pioneer works of E. Yablonovitch\cite{Yablonovitch_PRL_1987} and S. John\cite{John_PRL_1987}, and it remains an attractive research area (for  details about theoretical, experimental and numerical techniques, see Refs. \cite{Petritsch_Book_2017,Gong_Book_2014,Laine_Book_2010,Sukhoivanov_Book_2009,Joannopoulos_Book_2008,Sakoda_Book_2004}). In these structures the propagation of light can be controlled through a periodic modulation of the dielectric constant, which is analogous to the propagation of electrons in crystals, with the photon being responsible by the propagation and processing of information along the system \cite{Markos_Book_2008}. PCs are currently used in many technological applications \cite{Petritsch_Book_2017,Gong_Book_2014,Laine_Book_2010} such as waveguides, optical fibers, optical computing devices, lasers, and solar cells.

Often hailed as a wonder material due to its impressive physical properties, graphene has opened several venues of basic science exploration and it is a material that has a tremendous technological potential \cite{Novoselov_Science_2004}. In graphene, carbon atoms with $sp^{2}$ hybridization are strongly and densely attached creating a planar hexagonal crystalline lattice wich makes it a material that has the exotic property of being a two-dimensional arrangement with thickness of a single atom\cite{Wolf_Book_2016,Torres_Book_2014}. The honeycomb lattice structure of graphene and its two sublattices are responsible for a variety of novel physical phenomena\cite{Wolf_Book_2014,Aoki_Book_2013}. Among the unique physical properties of graphene\cite{Rao_Book_2013}, we highlight lightness and rigidity, high thermal\cite{Pereira2013,Xu2014} and electrical conductivity\cite{Novoselov_Science_2004}. Graphene has been considered for potential technological applications in telecommunication, flexible displays, batteries and in the production of electronic devices with good heat dissipation\cite{Sharon_Book_2015,Souza_Book_2014,Warner_Book_2012}. Very recently several research groups have been investigating materials composed of photonic crystals and graphene, giving rise to a new research area: \textit{graphene nanophotonics}\cite{Nikitin_JO_2013,Abajo_Science_2013}. In particular, it is has been shown that embedding graphene between adjacent layers of a periodic dielectric multilayer, allows tailoring photonic band gaps in the dispersion relation of the structure\cite{Soukoulis_PRB_2013}.

After the discovery of quasicrystals by Shechtman and co-workers in 1984\cite{Shechtman_PRL_1984}, a great interest in the so-called quasiperiodic disorder has aroused\cite{Limonov_Book_2016,Fujiwara_Book_2014,Negro_Book_2013,Suck_Book_2010,Barber_Book_2008,Janssen_Book_2008}. Because of his work on quasicrystals, which created a wide field of research in condensed matter, he was awarded the Nobel Prize for Chemistry in 2011\cite{Shechtman_Nobel_2011}. One of the most important reasons for that is because quasiperiodic systems can be defined as an intermediate state between an ordered crystal and a disordered solid. On the theoretical side, a wide variety of particles, namely, electrons, phonons, plasmon-polaritons, and magnons, have been studied in quasiperiodic systems\cite{Albuquerque_Book_2004}. On the experimental side, in a pioneer work in the 90’s, Munzar and collaborators studied GaAs/GaAlAs Fibonacci superlattices. Those authors theoretically and experimentally studied the reflectance and electronic multifractal spectra of the system\cite{Munzar}. In particular, multilayered dielectric structures arranged in a quasiperiodic fashion are called photonic quasicrystals\cite{Vardeny_NP_2013} (PQCs), which present a fundamental role for the next generation of optical devices\cite{Macia_RPP_2006}.

In this work we investigate the propagation of photons in dieletric multilayers, arranged according to the Fibonacci quasiperiodic sequence, with embedded graphene sheets. The aim of this work is twofold: we investigate (i) the competition between graphene induced band gaps and those ones produced by Bragg reflections and (ii) the effects of long-range correlation induced by the quasiperiodic disorder. This paper is organized as follows. In Section 2 we discuss the theoretical model with emphasis on the description of the Fibonacci quasiperiodic sequence and the transfer matrix technique, which is employed to simplify the algebra that can  otherwise be quite involved. The numerical results for dispersion relation and transmittance, illustrating the band gaps, are discussed in Section 3. Section 4 is devoted to discuss the physical origin of the graphene induced band gaps. Finally, our findings are summarized in Section 5.

\section{Physical model}\label{sec:model}

\begin{figure}[h]
  \centering
  \includegraphics[height=.4\textheight]{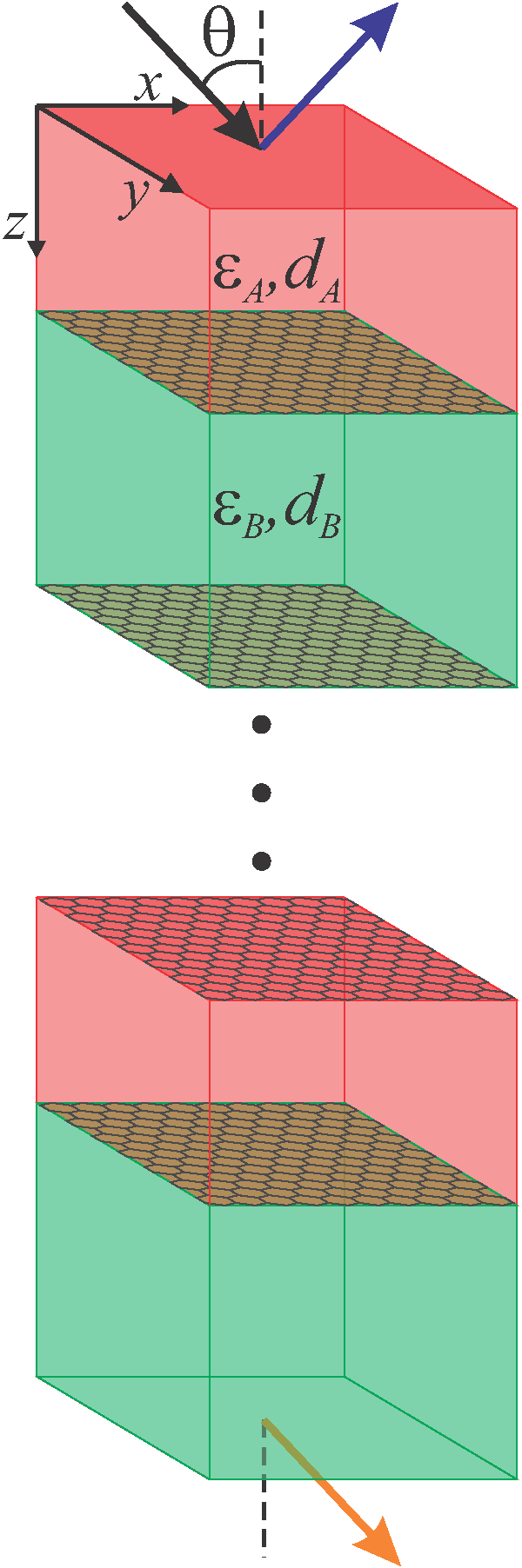}
  \caption{Structure of the one-dimensional photonic crystal composed by alternating dieletric layers $A$ and $B$. The graphene monolayers are embedded between two consecutive layers.}\label{fig:1D-GPC}
\end{figure}

Our model describes two isotropic non-magnetic dielectric materials $A$ and $B$, with permittivity $\epsilon_ {A}$ and $\epsilon_ {B}$ and thickness $d_ {A}$ and $d_ {B}$, respectively, as described in Fig.\ \ref{fig:1D-GPC}. The graphene layers, embedded between adjacent dielectric layers, present a frequency-dependent conductivity \cite{Madani_PhysicaB_2013,Ning_JO_2014,Farhat_OE_2013},
\begin{equation}\label{eq:sigma}
\displaystyle	\sigma_{g}(\omega)=\sigma_{g}^{intra}(\omega)+\sigma_{g}^{inter}(\omega),
\end{equation}
with

\begin{equation}\label{eq:sigma_intra}
\displaystyle	\sigma_{g}^{intra}(\omega)=i\frac{e^{2}}{\pi\hbar\left(\hbar\omega+i\Gamma\right)}\left\{\mu_{c}+2k_{B}T_{K}\ln\left[e^{\left(-\mu_{c}/k_{B}T_{K}\right)}+1 \right] \right\},
\end{equation}

and

\begin{equation}\label{eq:sigma_inter}
\displaystyle	\sigma_{g}^{inter}(\omega)=i\frac{e^{2}}{4\pi\hbar}\ln\left[ \frac{2\left|\mu_{c}\right|-\left(\hbar\omega+i\Gamma\right)}{2\left|\mu_{c}\right|+\left(\hbar\omega+i\Gamma\right)} \right].
\end{equation}

\noindent Here $e$ is the electronic charge, $\hbar=h/2\pi$ is the reduced Planck's constant, $k_{B}$ is the Boltzmann's constant, $T_{K}$ is the temperature in Kelvin, $\Gamma$ the damping constant of graphene and $\mu_{c}$ is the chemical potential, which can be controlled by the gate voltage. The intraband contribution is due to scattering from phonons, electrons and impurities, while the interband contribution is due to electron-hole recombination. In particular $\hbar\omega$ is much smaller than the chosen chemical potential, so that the interband contribution is neglectable\cite{Aliofkhazraei_Book_2016}.

\begin{figure}[h]
  \centering
  \includegraphics[width=.4\textwidth]{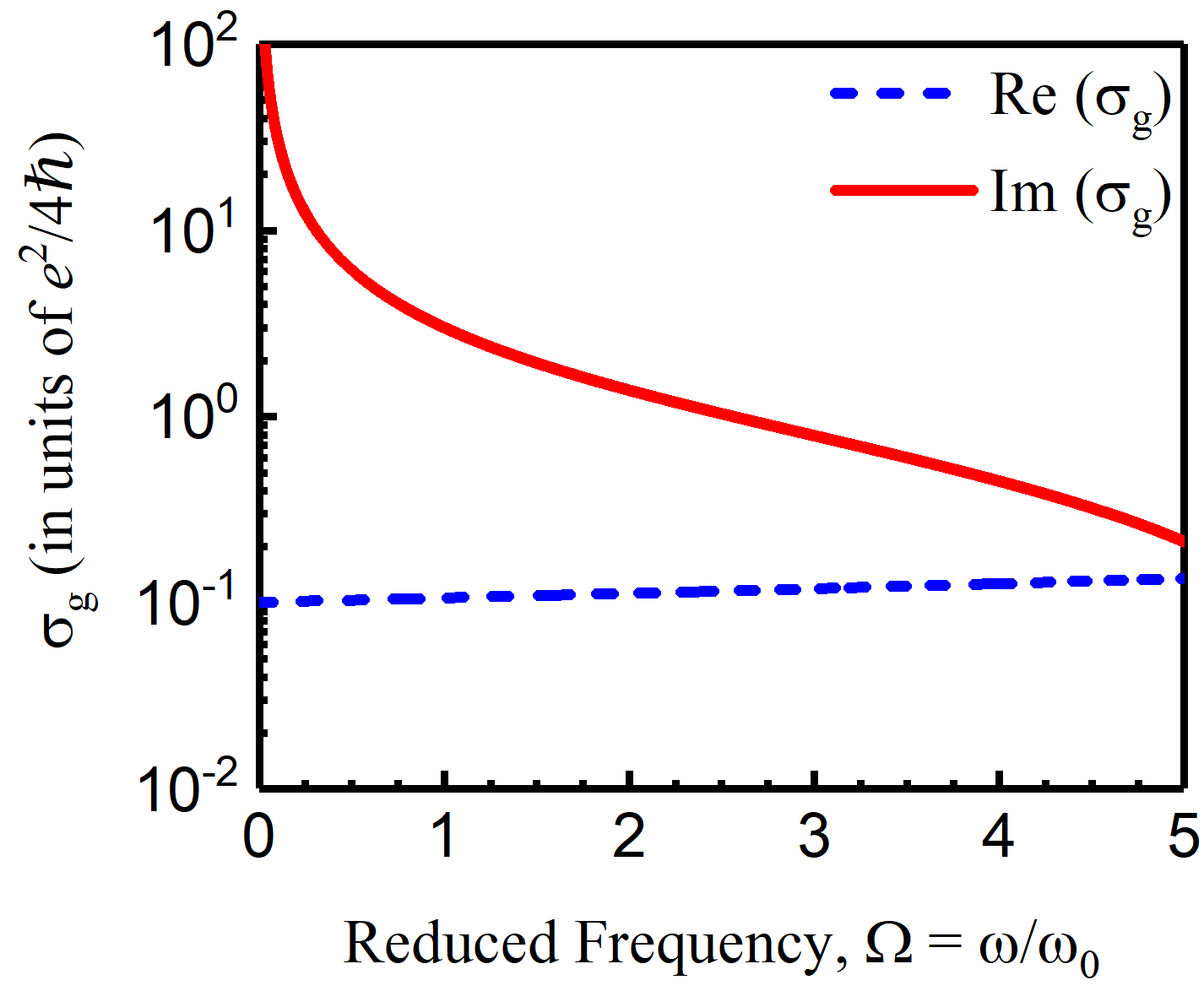}
  \caption{Surface conductivity $\sigma_{g}$ (in units of $e^{2}/4\hbar$) of graphene as a function of the reduced frequency $\Omega=\omega/\omega_{0}$. The dashed blue (solid red) line represents the real (imaginary) part of $\sigma_{g}$. From the figure, we can expect effects of graphene on the light propagation in the range $\Omega\leq 1$.}\label{fig:SIGMA}
\end{figure}

In Fig. \ref{fig:SIGMA} we plot the real and imaginary parts of $\sigma_{g}$ as a function of reduced frequency, $\Omega=\omega/\omega_{0}$, with $\omega_{0}=2\pi c/\lambda_{0}\approx 31,4$ THz, for a central wavelength $\lambda_{0}=60$ $\mu$m. The real part does not have a strong dependence on the frequency, but the imaginary part shows a strong dependence in the low-frequency region. As we will show later, the imaginary part of $\sigma_{g}$ is directly associated with reflection of electromagnetic waves.

The multilayers considered in this work are composed by two dieletric building blocks $A$ and $B$, which are arranged according to the quasiperiodic Fibonacci sequence\cite{Costa_JPCM_2013}. The $n$-th generation of the Fibonacci sequence can be obtained by appending the $(n-2)$-th generation to the $(n-1)$-th one, i.e., 
\begin{equation}\label{eq:sequencia_fibonacci1}
	S_{n}=S_{n-1}S_{n-2}\qquad (\textrm{with $n\geq 2$}).
\end{equation}
This algorithm construction requires  initial  conditions  which  are  chosen  to  be $S_{0}=B$ e $S_{1}=A$. The  Fibonacci  generations  can  also  be  alternatively obtained  by  an  iterative  process  from  the  substitution rules (or inflation rules) $A \rightarrow AB$ and $B \rightarrow A$, in such way that the first Fibonacci generations are
\begin{eqnarray*}\label{eq:geracoes_sequencia_fibonacci}
	& & S_{0}=[B], \hspace {0.25cm} S_{1}=[A], \hspace {0.25cm} S_{2}=[A|B], \hspace {0.25cm} S_{3}=[A|B|A],\\ \nonumber
	& & S_{4}=[A|B|A|A|B], \hspace{0.25cm} S_{5}=[A|B|A|A|B|A|B|A].
\end{eqnarray*}

\noindent In a given Fibonacci generation $S_n$, the total number of building blocks is given by the Fibonacci number $F_{n}$, which is obtained by the relation
\begin{equation}\label{eq:numero_fibonacci}
	F_{n}=F_{n-1}+F_{n-2} \quad (\textrm{$n\geq 2$}),
\end{equation}
with $F_{0}=F_{1}=1$. As the generation index increases ($n\gg1$), the ratio $F_n /F_{n-1}$ approaches $\phi$, an irrational number known as the Golden Mean, such that
\begin{equation}\label{eq:sigma_sequencia_fibonacci}
	\phi=\lim_{n\rightarrow \infty} \frac{F_{n}}{F_{n-1}}=\frac{1 + \sqrt{5}}{2}.
\end{equation}

For the photonic quasicrystals considered here, whose unit cell is given by the Fibonacci generation $S_n$, repeated $N$ times, the transfer matrix $M_n$ which relates the amplitude of the electromagnetic wave at the interfaces is given by\cite{Zaleki_JNanophoton_2016,Madani_PhysicaB_2013,Zhan_JPCM_2013}
\begin{equation}\label{eq:tsn}
	\displaystyle M_{n} = \left(M_{A}M_{B}\cdots M_{B}M_{A}\right)^{N}=\left(\prod M_{j}\right)^{N}
\end{equation}
with
\begin{eqnarray}
	\displaystyle M^{TE}_{j}(d_{j},\omega) & = & \left[
	\begin{array}{c}
		\cos(k_{zj}d_{j}) \\
		\sigma_{g}\cos(k_{zj}d_{j})+iq_{j}\sin(k_{zj}d_{j})\nonumber
	\end{array}
	\right.\\
	& & \left.
	\begin{array}{c}
		\left(\frac{i}{q_{j}}\right)\sin(k_{zj}d_{j}) \\
		\cos(k_{zj}d_{j})+\left(\frac{i\sigma_{g}}{q_{j}}\right)\sin(k_{zj}d_{j})
	\end{array}
	\right]
\end{eqnarray}
and
\begin{eqnarray}
	\displaystyle M^{TM}_{j}(d_{j},\omega) & = & \left[
	\begin{array}{c}
		\cos(k_{zj}d_{j})-i\sigma_{g}q_{j}\sin(k_{zj}d_{j}) \\
		iq_{j}\sin(k_{zj}d_{j}) \nonumber
	\end{array}
	\right.\\
	& & \left.
	\begin{array}{cc}
		-\sigma_{g}\cos(k_{zj}d_{j})+\left(\frac{i}{q_{j}}\right)\sin(k_{zj}d_{j}) \\
		\cos(k_{zj}d_{j})
	\end{array}
	\right].
\end{eqnarray}
Here $q_{j}=-\frac{k_{zj}}{\omega\mu_{0}}$ (for TE waves) and $q_{j}=\frac{k_{zj}}{\omega\epsilon_{0}\epsilon_{j}}$ (for TM waves). Also, $k_{zj}$ is the $z$-component of the wave vector within medium $j$ ($j=A$ or $B$), which is given by\cite{Brandao_OM_2015}
\begin{eqnarray}
	\displaystyle k_{zj} = \left\{
	\begin{array}{c}
		\textit{\textcolor{white}{i}}\left[ \epsilon_{j}(\omega/c)^{2} - k^{2}_{x0} \right]^{1/2}\quad (\epsilon_{j}(\omega/c)^{2}\geq k^{2}_{x0}), \\
		\\
		i\left[ k^{2}_{x0} - \epsilon_{j}(\omega/c)^{2} \right]^{1/2}\quad (\epsilon_{j}(\omega/c)^{2}< k^{2}_{x0}),
	\end{array}
	\right.
\end{eqnarray}
where $\epsilon_{j}$ is the dielectric constant of medium $j$, $c$ is the speed of light in vacuum and $k_{x0}$ is the $x$-component of the incoming wave vector.

The coefficients of transmission $T$, reflection $R$ and absorption $A$ are obtained from the elements of the transfer matrix $M_ {n}$, corresponding to the $n$-th generation of the Fibonacci sequence, and are given by\cite{Madani_PhysicaB_2013}
\begin{equation}\label{eq:transmissao}
	T=\left|\frac{2q_{0}}{q_{t}M_{11}+q_{0}M_{22}-M_{21}+q_{0}q_{t}M_{12}}\right|^{2},
\end{equation}
\begin{equation}\label{eq:reflexao}
	R=\left|\frac{q_{t}M_{11}-q_{0}M_{22}-M_{21}+q_{0}q_{t}M_{12}}{q_{t}M_{11}+q_{0}M_{22}-M_{21}+q_{0}q_{t}M_{12}}\right|^{2},
	\end{equation}
and
\begin{equation}\label{eq:absorcao}
	A=1-T-R.
\end{equation}
Here $M_{ij}$ are the elements of the transfer matrix $M_n$ and $q_{0}$, $q_{t}$ are the $q$ parameters of the incoming medium and outgoing medium, respectively.

The dispersion relation of the eletromagnetic waves propagating in the quasiperiodic structure is given by \cite{Markos_Book_2008}
\begin{equation}\label{eq:relacao_de_dispersao}
	\cos(QD)=\left (\frac{1}{2} \right )\textrm{Tr}\left[M_{n}\right],
\end{equation}
where $Q$ is the Bloch wave vector, $D=N\times D_{n}$ is the size of the multilayer ($D_{n}$ is the unit cell size of the $n$-th Fibonacci generation) and $\textrm{Tr}\left[M_{n}\right]$ is the trace of the transfer matrix $M_n$.

\section{Numerical Results}\label{sec:Numerical Results}
Let us now present our numerical results to illustrate the optical transmission spectra and dispersion relation of the Fibonacci quasiperiodic photonic crystal. We consider the same parameters used in Ref.\ \cite{Brandao_OM_2015}, i.e., those appropriate for silicon dioxide (building block $A=\textrm{SiO}_{2}$) and titanium dioxide (building block $B=\textrm{TiO}_{2}$). We also consider the individual layers as quarter-wave layers, for which the quasiperiodicity is expected to be more effective \cite{Vasconcelos_PRB_1999}, with central wavelength $\lambda_{0}=60$ $\mu$m. These conditions yield a thickness $d_{j}=60/4n_{j}$ $\mu$m ($j=A$ or $B$), such that $n_{A}d_{A}=n_{B}d_{B}$, providing $d_{A}\approx 10.34$ $\mu$m e $d_{B}\approx 6.52$ $\mu$m. Their refractive index around the central wavelength $\lambda_{0}$ are $n_{A}=\sqrt{\epsilon_{A}}=1.45$ and $n_{B}=\sqrt{\epsilon_{B}}=2.30$, respectively. We consider the photonic quasicrystal surrounded by vacuum, such that $\epsilon_{0}=\epsilon_{t}=1$ at $q_{0}$ and $q_{t}$. Here, we assume the surface conductivity of graphene given by Eq. \ref{eq:sigma} with $\mu_{c}=0.2$ eV, $\Gamma=0$ eV and $T_{K}=300$ K.

\subsection{Transmittance and dispersion relation for normal incidence ($\theta=0^{\circ}$)}\label{subsec:TxOmega_0}

\begin{figure}[h]
  \centering
  \includegraphics[width=.5\textwidth]{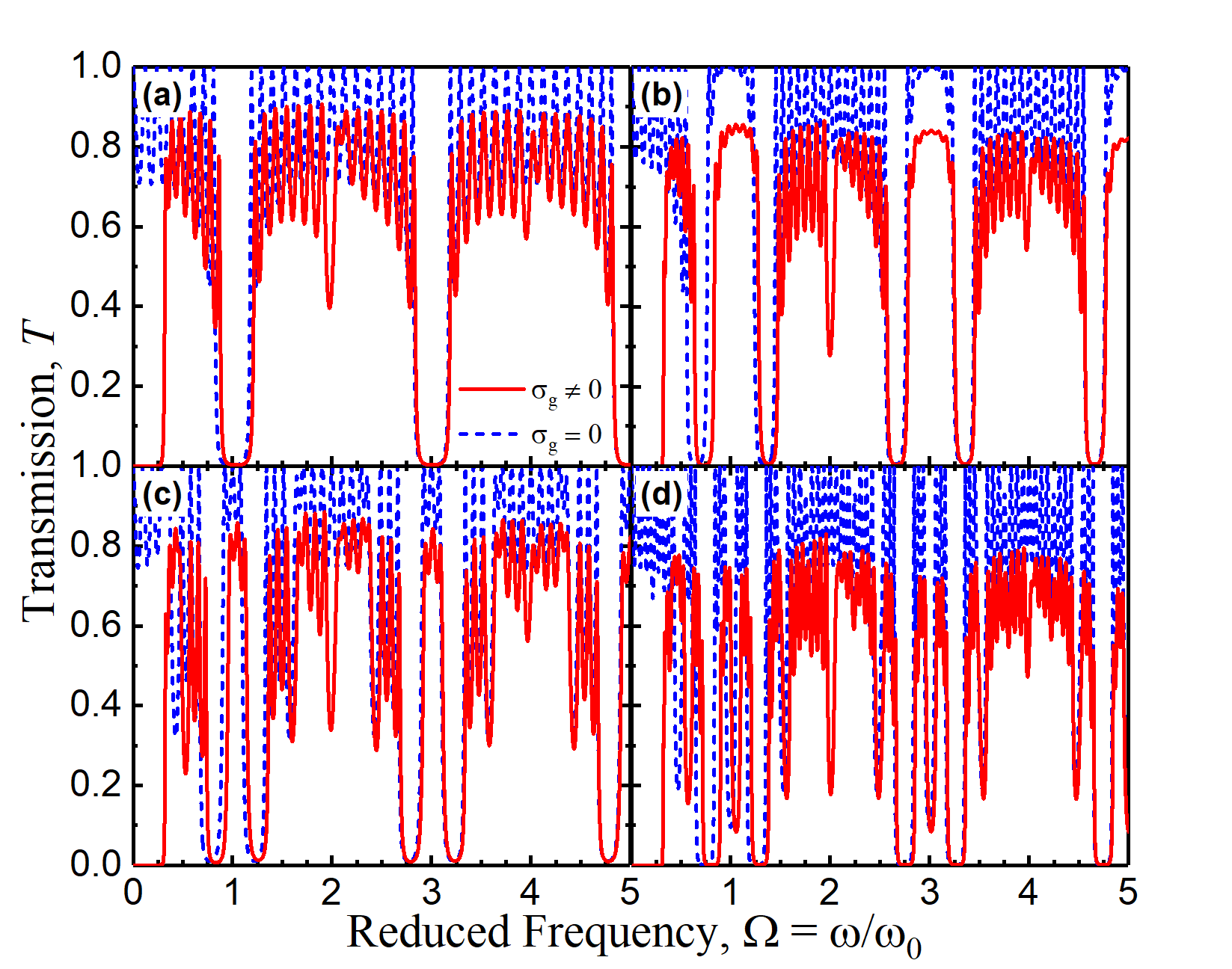}
  \caption{Transmission spectra as a function of the reduced frequency $\Omega=\omega/\omega_{0}$ for normal incidence with (red solid line) and without (blue dashed line) graphene monolayers at the interfaces.  (a) 2nd generation, with $N=4$ ($[A|B]^{4}$), (b) 3rd generation, with $N=4$ ($[A|B|A]^{4}$), (c) for 4th generation, with $N=3$ ($[A|B|A|A|B]^{3}$), and (d) for 5th generation, with $N=3$ ($[A|B|A|A|B|A|B|A]^{3}$). The presence of the graphene between the dielectric layers reduces the transmissivity on the whole frequency range. Moreover, it induces a transmission gap in the low-frequency region ($\Omega \leq 0.38$).}
\end{figure}

In Fig.\ 3 we show the transmittance versus reduced frequency $\Omega=\omega/\omega_{0}$, with $\omega_{0}=2\pi c/\lambda_{0}$, for normal incidence, with (red solid lines) and without (blue dashed lines) graphene at interfaces.  Panel (a) is the 2nd Fibonacci generation (the periodic case) with $N=4$ ($4\times F_{2}=8$ building blocks), panel (b) corresponds to the 3rd Fibonacci generation with $N=4$ ($4\times F_{3}=12$ building blocks), panel (c) is the 4th Fibonacci generation with $N=3$ ($3\times F_{4}=15$ building blocks) and panel (d) corresponds to the 5th Fibonacci generation with $N=3$ ($3\times F_{5}=24$ building blocks). We observe from Fig.\ 3 that the presence of the graphene monolayers between the dielectric layers reduces the transmissivity on the whole range of frequency. Moreover, it induces a transmission gap in the low-frequency region ($\Omega \leq 0.38$). On the other hand, for $\Omega> 0.38$ the position or width of the transmission gaps are not substantially modified by the presence of the graphene monolayers. This features can be explained, according to Fig.\ 2, because the effects of the surface conductivity of graphene are expected to be more effective in the low frequency region. From Fig.\ 3 we conclude that the graphene induced band gap does not depend on the Fibonacci generation index $n$.

\begin{figure}[h]
  \centering
  \includegraphics[width=.5\textwidth]{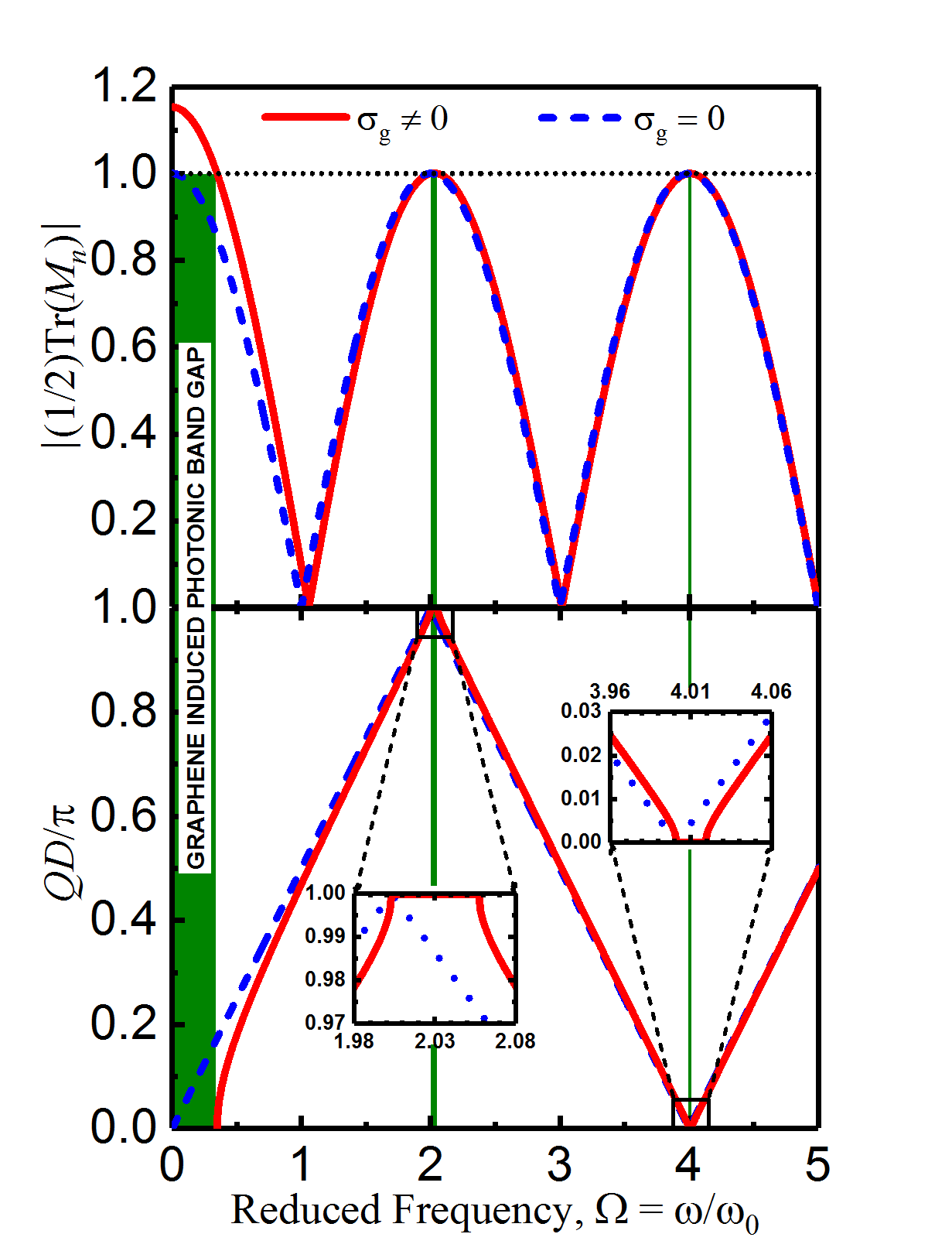}
  \caption{Plots of $|(1/2)\textrm{Tr}\left[M_n\right]|$ (top) and dimensionless Bloch's wave vector $QD/\pi$ (bottom) versus the reduced frequency $\Omega=\omega/\omega_{0}$ for the 1st Fibonacci generation with graphene at the interfaces and $N=10$ ($[A]^{10}$).}\label{fig:DISPERSION-TRACE-1GER}
\end{figure}

Fig.\ 4 shows the plots of $|(1/2)\textrm{Tr}\left[M_n\right]|$ (on top) and dimensionless Bloch wave vector $QD/\pi$ (on bottom), versus the reduced frequency $\Omega=\omega/\omega_{0}$, for the 1st Fibonacci generation (SiO$_2$) with $N=10$ ($10\times F_{1}=10$ building blocks), with (red solid line) and without (blue dashed line) graphene monolayers. We observe  the presence of photonic band gaps from the insets of Fig.\ 4 (green stripes in the figure) which are not present in the absence of graphene, as expected. Therefore, these photonic band gaps are exclusively due to the presence of the graphene monolayers at the interfaces. These \textit{graphene induced photonic band gaps} (GIPBG) are unsual because they do not come from Bragg reflections. We observe three GIPBGs: the first one is wider, for $\Omega\le 0.38$, and the other two are around $\Omega\sim 2.0$ and $\Omega\sim 4.0$, respectively. The low-frequency GIPBG is wider because the effects of graphene are expected to be more pronounced in the low-frequency region as discussed in Sec. II (and shown in Fig. \ref{fig:SIGMA}).

\begin{figure}[h]
  \centering
  \includegraphics[width=.5\textwidth]{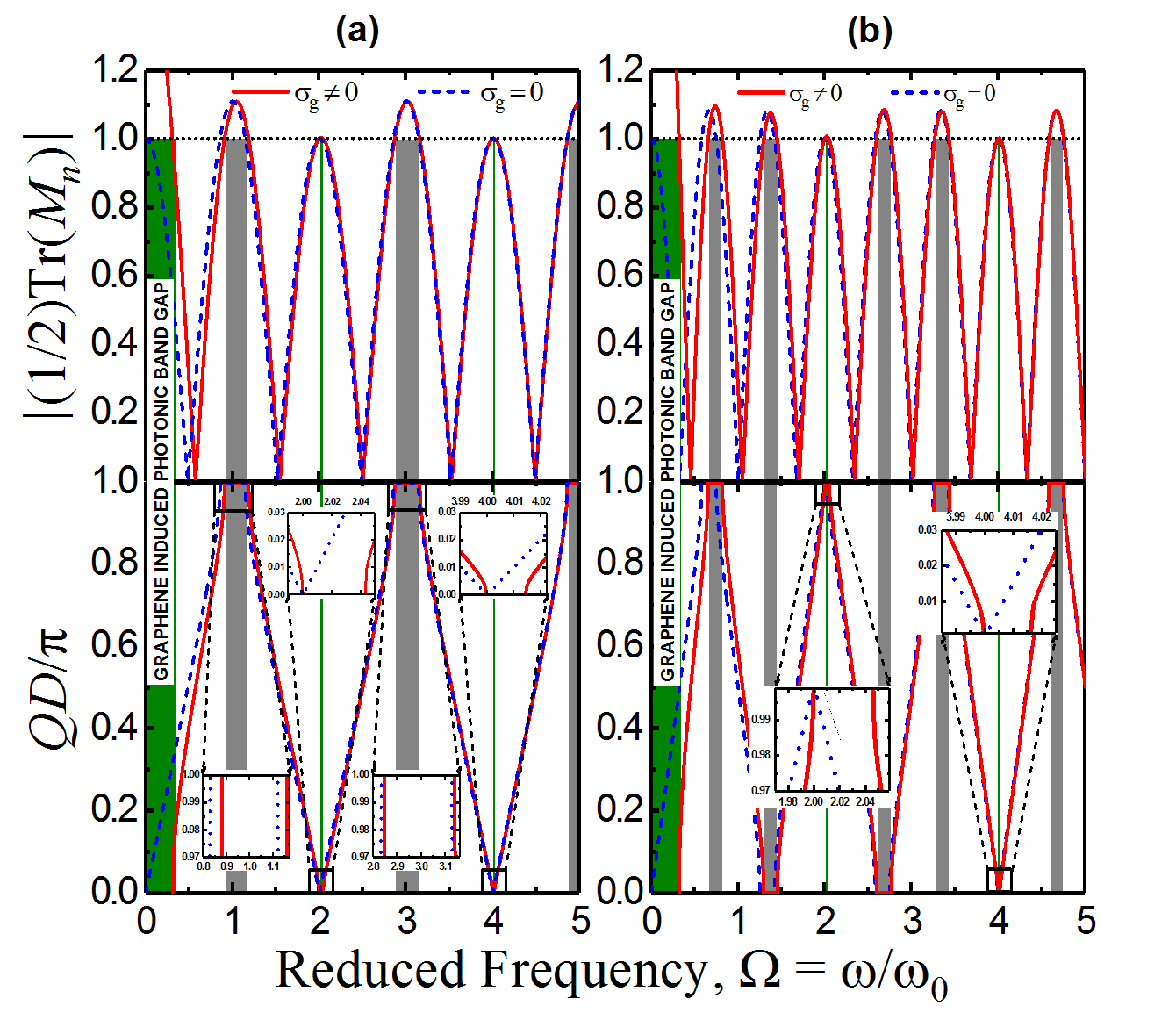}
  \caption{Same as Fig. \ref{fig:DISPERSION-TRACE-1GER} for the (a) 2nd and (b) 3rd generations of the Fibonacci sequence, with $N=4$ for both cases.}\label{fig:DISPERSION-TRACE-23GER}
\end{figure}

In Fig.\ 5 we show the plots of $|(1/2)\textrm{Tr}\left[M_n\right]|$ (on top) and dimensionless Bloch wave vector $QD/\pi$ (on bottom), versus the reduced frequency $\Omega=\omega/\omega_{0}$, for (a) 2nd (which corresponds to the periodic photonic crystal) and (b) 3rd generations of the Fibonacci sequence, both cases with $N=4$, i.e., $4\times F_{2}=8$ building blocks and $4\times F_{3}=12$ building blocks, respectively. Now we observe ordinary gaps produced by Bragg reflections, which are slightly shifted to higher frequencies and are depicted by gray stripes in the figure. We also observe GIPBGs (green stripes in the figure): one wider for $\Omega\le 0.38$ and two narrower gaps. The insets in the figure show the GIPBGs in detail. The insets in Fig.\ 5b show only the gaps formed by graphene because they are narrower and difficult to visualize. The spectra for the 2nd and 3rd Fibonacci generations are qualitatively similar, except for the number of gaps, which is expected due to the different unit cell sizes. We should remark that the wider GIPBG for $\Omega\le 0.38$ is independent of  Fibonacci generation. Therefore, we conclude that the wider GIPBG is exclusively  due to the presence of the graphene monolayers at the interfaces, and  it is not related to the long range effects induced by the quasiperiodic substitutional sequence. 

\begin{figure}[h]
  \centering
  \includegraphics[width=.5\textwidth]{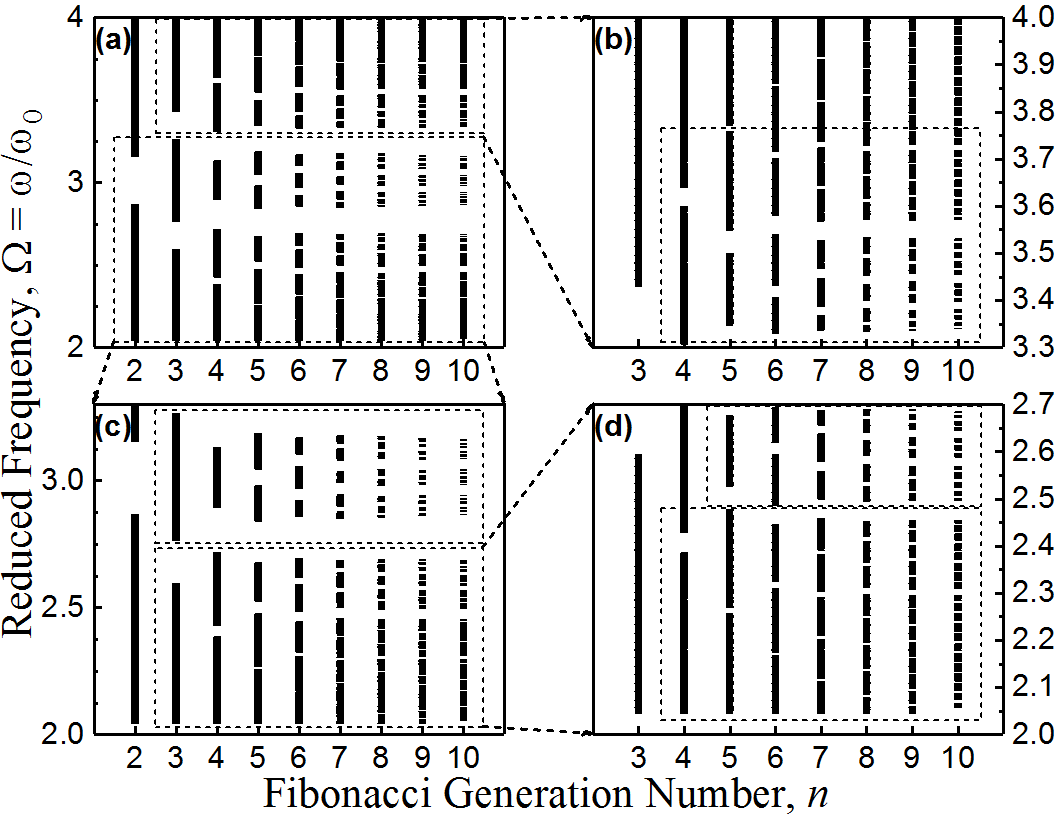}
  \caption{Distribution of the allowed photonic bandwidths for the Fibonacci photonic quasicrystal as a function of the Fibonacci generation index $n$.}\label{fig:OMEGAxN}
\end{figure}

In order to characterize the self-similar behavior of the photonic spectra, in Fig.\ 6 we present the allowed photonic bandwidths distribution versus the Fibonacci generation index $n$, for normal incidence. The insets show amplifications of some regions of the spectra. It is easy to observe that the insets reveal a self-similar behavior of the allowed photonic bandwidths. This is an expected behavior and it is considered the basic signature of quasiperiodic systems.

\subsection{Transmittance for oblique incidence ($\theta \neq 0$)}\label{subsec:TxOmega_0}\label{sec:Trilayer model}

\begin{figure}[h]
  \centering
  \includegraphics[width=.5\textwidth]{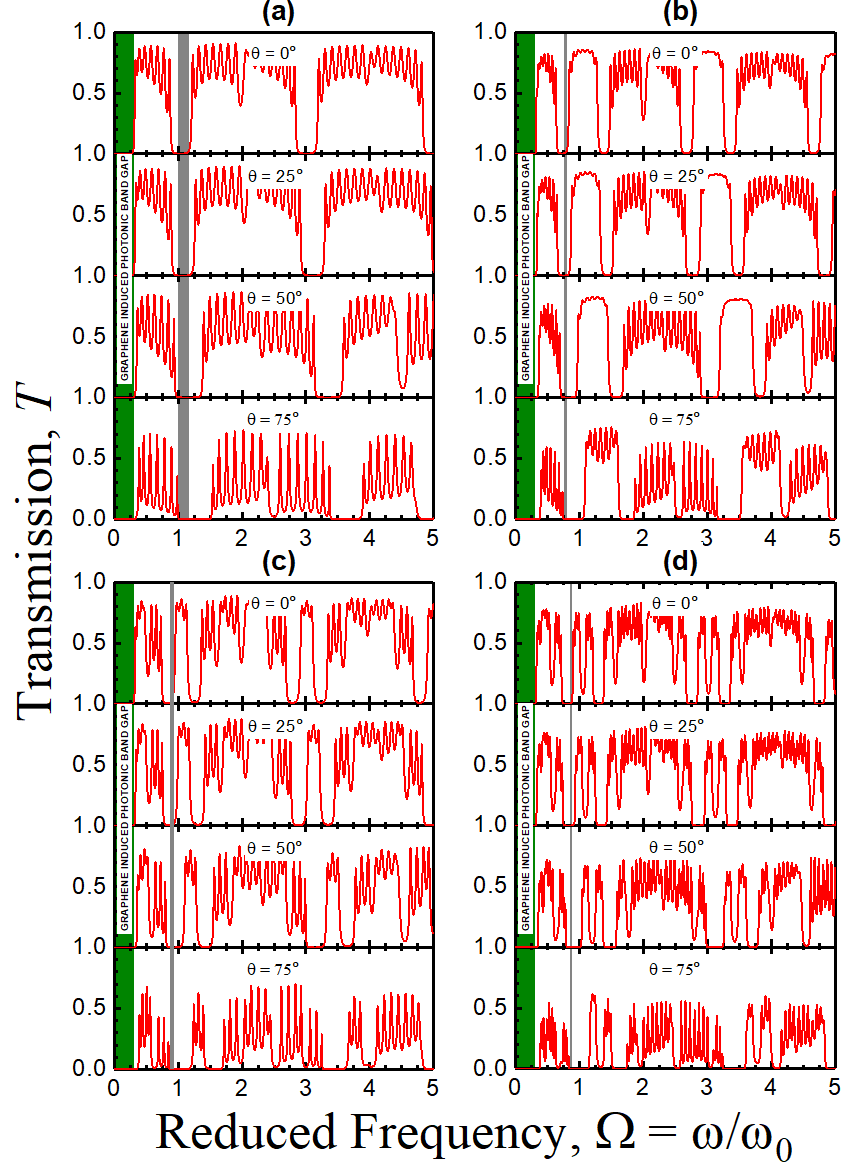}
  \caption{Transmission $T$ of TE waves as a function of the reduced frequency $\Omega=\omega/\omega_{0}$ for (a) 2nd generation, with $N=4$ ($[A|B]^{4}$), (b) 3rd generation, with $N=4$ ($[A|B|A]^{4}$), (c) for 4th generation, with $N=3$ ($[A|B|A|A|B]^{3}$), and (d) for 5th generation, with $N=3$ ($[A|B|A|A|B|A|B|A]^{3}$). }\label{fig:TxOMEGA_TE}
\end{figure}

Let us now consider the more general case of oblique incidence. In Fig.\ 7  is shown the transmissivity spectra for the Fibonacci photonic crystal, with embedded graphene, for some specific values of the angle of incidence ($\theta= 0^{\circ}$, $25^{\circ}$, $50^{\circ}$ and $75^{\circ}$), for 2nd, 3rd, 4th and 5th Fibonacci generations, considering transverse electric (TE) polarization. We observe that GIPBGs (green stripes in the figure) are also \textit{omnidirectional photonic band gaps}, i.e., the light cannot propagate in the structures for $\Omega\leq 0.38$ whatever is the angle of incidence. We also observe the usual band gaps (gray stripes in the figure), produced by Bragg reflections, which are also omnidirectional photonic band gaps. Regarding the  transmissivity,  the transmission coefficient decreases as the angle of incidence increases for all generations of the Fibonacci sequence. This is because the reflection coefficient, due to the presence of the graphene monolayers, increases. We remark that the graphene induced omnidirectional photonic band gap presents the same width. This ratifies that it is really independent of the Fibonacci generation index $n$. On the other hand, the usual Bragg omnidirectional photonic band gaps get narrower and narrower as the generation index $n$ increases.

\begin{figure}[h]
  \centering
  \includegraphics[width=.5\textwidth]{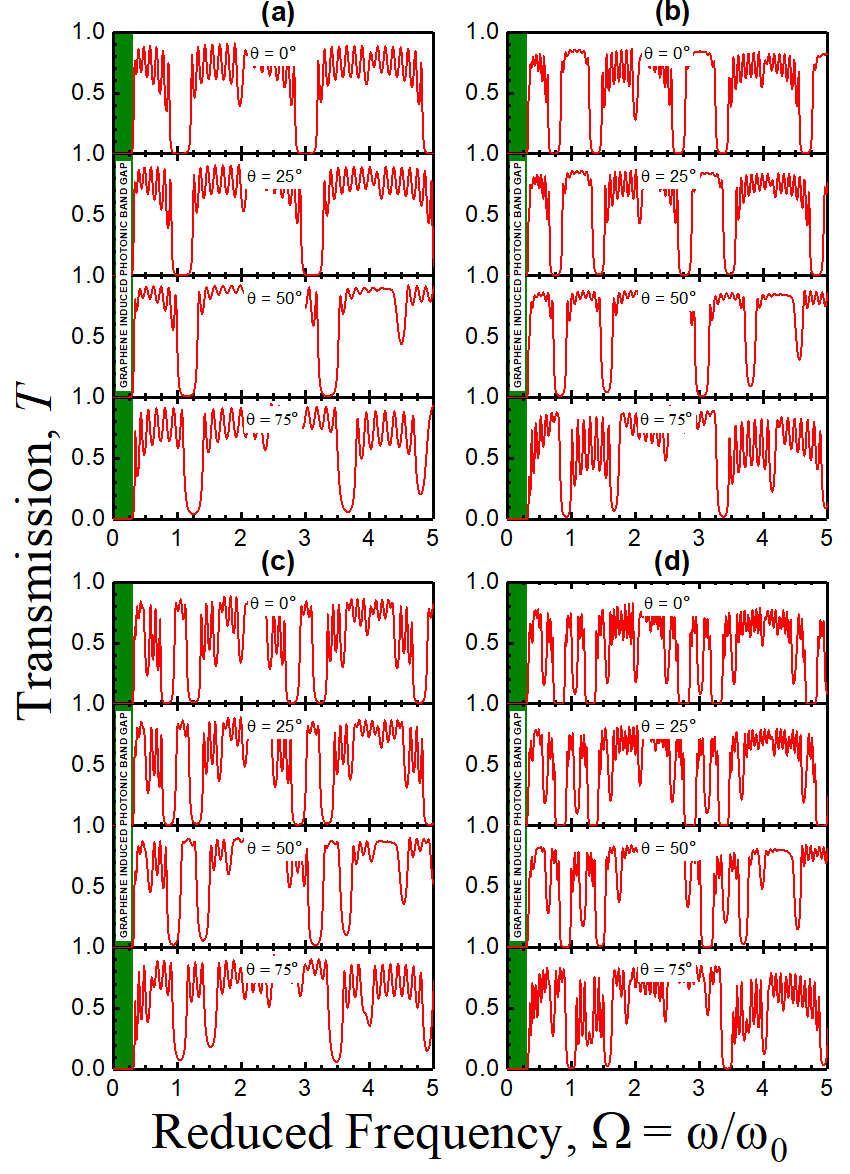}
  \caption{ Transmission $T$ of TM waves as a function of the reduced frequency $\Omega=\omega/\omega_{0}$ for (a) 2nd generation, with $N=4$ ($[A|B]^{4}$), (b) 3rd generation, with $N=4$ ($[A|B|A]^{4}$), (c) for 4th generation, with $N=3$ ($[A|B|A|A|B]^{3}$), and (d) for 5th generation, with $N=3$ ($[A|B|A|A|B|A|B|A]^{3}$),. }\label{fig:TxOMEGA_TM}
\end{figure}

Now, we consider the transverse magnetic (TM) polarization case with oblique incidence. We consider the same Fibonacci generations ($n=2,3,4$ and $5$) and angles of incidence ($\theta= 0^{\circ}$, $25^{\circ}$, $50^{\circ}$ and $75^{\circ}$) of the TE case. The corresponding transmissivity spectra is illustrated in Fig.\ 8. We can observe, similar to the TE case,  the presence of GIPBGs (green stripes in the figure), which are \textit{omnidirectional photonic band gaps}, for $\Omega\leq 0.38$. We can conclude that the GIPBGs \textit{do not depend} of the polarization of light. Moreover, the usual Bragg omnidirectional photonic band gaps \textit{are not present}. It is clear from Fig.\ 8 that, unlike what happens with TE polarization, the angle of incidence has a minor influence on the transmission coefficient. 

\subsection{Transmission spectra as a function of the reduced frequency ($\Omega$) and angle of incidence ($\theta$)}\label{sec:Dynamic properties}

\begin{figure}[h]
  \centering
  \includegraphics[width=.5\textwidth]{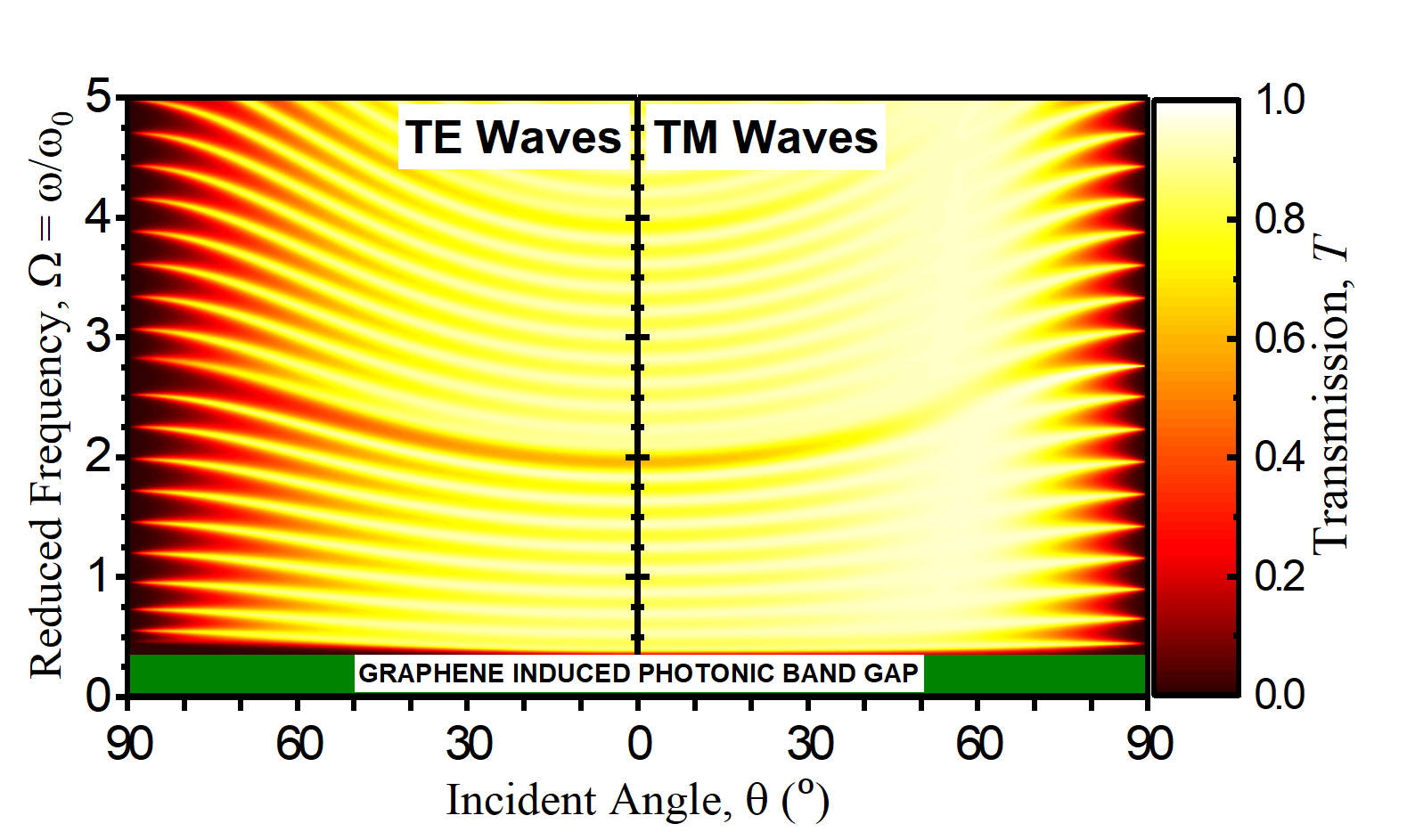}
  \caption{Transmission spectra as a function of the reduced frequency $\Omega$ and angle of incidence $\theta$, for TE (left) and TM (right) waves for the 1st generation of the Fibonacci sequence with graphene at the interfaces and $N=10$ ($[A]^{10}$). }\label{fig:1GER-M=10-3D }
\end{figure}

In Fig.\ 9 we show the transmittance coefficient, as a function of the angle of incidence and reduced frequency, for the 1st generation of the Fibonacci sequence with $N=10$ ($10\times F_{1}=10$ building blocks).  One can see in the low frequency region the GIPBG (green stripe), which is fully independent of the incidence angle $\theta $.   Regarding the transmission spectra, in Fig.\ 9 the black (white) color means a transmission coefficient equals to 0 (1). We notice that the spectra for TM waves is ``brighter'' than the spectra for the TE waves, which means that the transmission coefficient is higher for the former than for the latter. We can also note that the GIPBG is independent of the polarization of light (TE or TM), as mentioned above.

\begin{figure}[h]
  \centering
  \includegraphics[width=.5\textwidth]{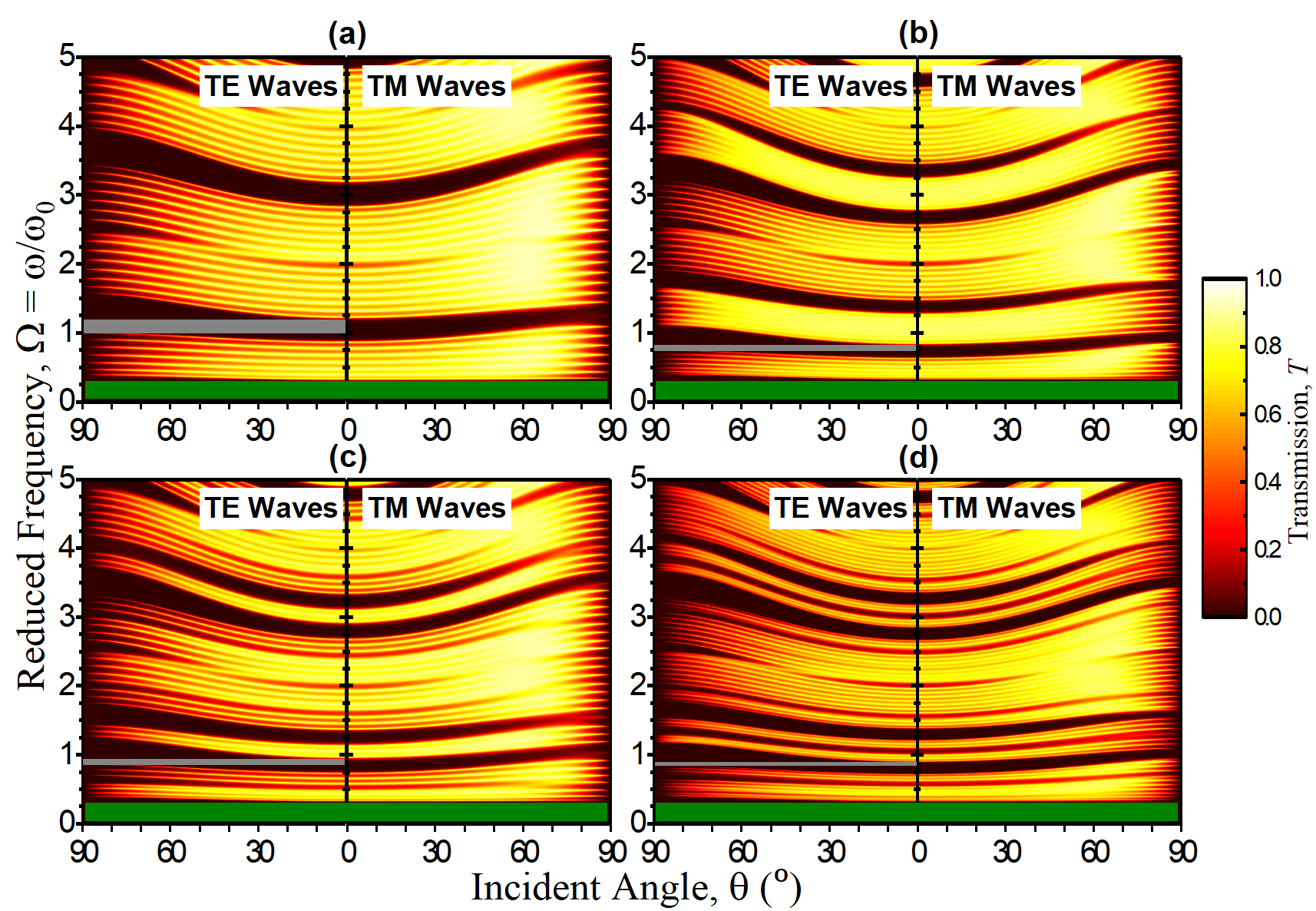}
  \caption{Transmission spectra as a function of the reduced frequency $\Omega$ and incident angle $\theta$ for TE (left) and TM (right) waves for: (a) second and (b) third, (c) fourth and (d) fifth generations of the Fibonacci sequence. The number of building blocks is the same as in Figs. 3, 7 and 8. }
  \label{fig:OMEGAxTHETAxT-2345GER}
\end{figure}

The transmission spectra as a function of the reduced frequency $\Omega$ and incident angle $\theta$, for TE and TM waves, for the second, third, fourth and fifth generations of the Fibonacci sequence are shown in Fig.\ 10. As in Fig.\ 9, the black (white) color means a transmission coefficient equal 0 (1). One can observe the presence of omnidirectional GIPBGs (green stripes) and Bragg photonic band gaps (gray stripes). Once again, our results show that GIPBGs are independent of the Fibonacci generation index $n$. On the other hand, the Bragg photonic band gaps get narrower and narrower, as the generation index increases. The GIPBGs emerge for both TM and TE cases, while the Bragg omnidirectional photonic band gaps emerge only for the TE case. For all Fibonacci generations  the spectra for TM waves is ``brighter'' than the spectra for the TE waves, which means that the transmission coefficient is higher for the TM case than for the TE one, as mentioned above.

\subsection{Transmission spectra as a function of the reduced frequency $\Omega$ and chemical potential $\mu_{c}$, for normal incidence ($\theta=0^{\circ}$)}\label{sec:Dynamic properties}

\begin{figure}[h]
  \centering
  \includegraphics[width=.5\textwidth]{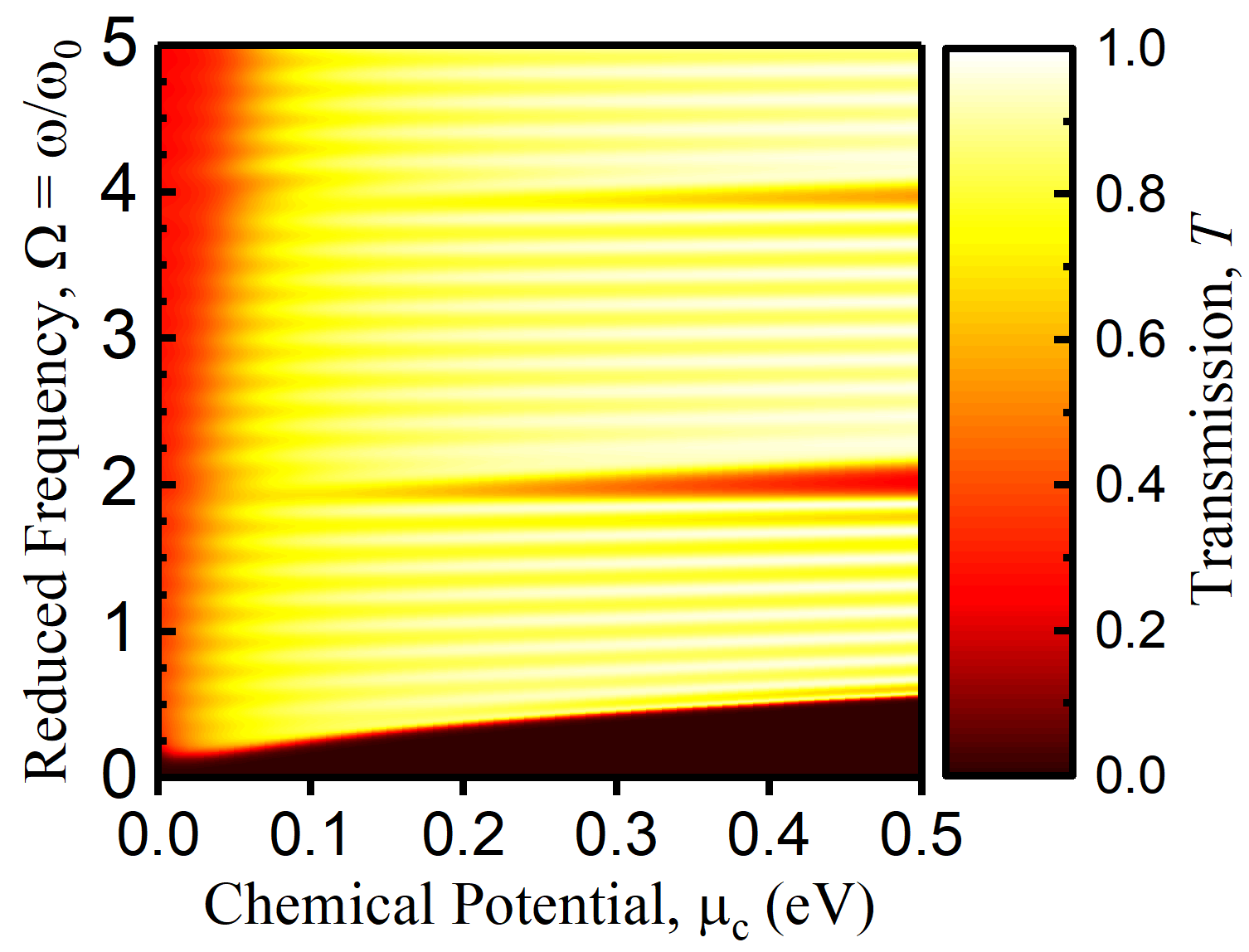}
  \caption{Transmission spectra as a function of the reduced frequency $\Omega=\omega/\omega_{0}$ and chemical potential $\mu_{c}$, normal incidence ($\theta=0^{\circ}$), for the 1st generation of the Fibonacci sequence with $N=10$ ($[A]^{10}$). }\label{fig:1GER-M=10-3D_MU}
\end{figure}

Last but not least, we consider the transmission spectra as a function of the reduced frequency $\Omega$ and chemical potential $\mu_{c}$, with normal incidence ($\theta=0^{\circ}$). For normal incidence the transmission spectra is exactly the same for both TE and TM cases. Fig.\ \ref{fig:1GER-M=10-3D_MU} shows the 1st generation of the Fibonacci sequence. As before the black (white) color means a transmission coefficient equal 0 (1). There is only one photonic band gap in the low frequency region (GIPBG). It is interesting to notice from Fig.\ 11 that we can adjust the width of the photonic band gap in a quasiperiodic photonic crystal by tuning of the chemical potential via a gate voltage. In fact, the width of the GIPBG monotonically increases as the chemical potential $\mu_{c}$ increases.

\begin{figure}[h]
  \centering
  \includegraphics[width=.5\textwidth]{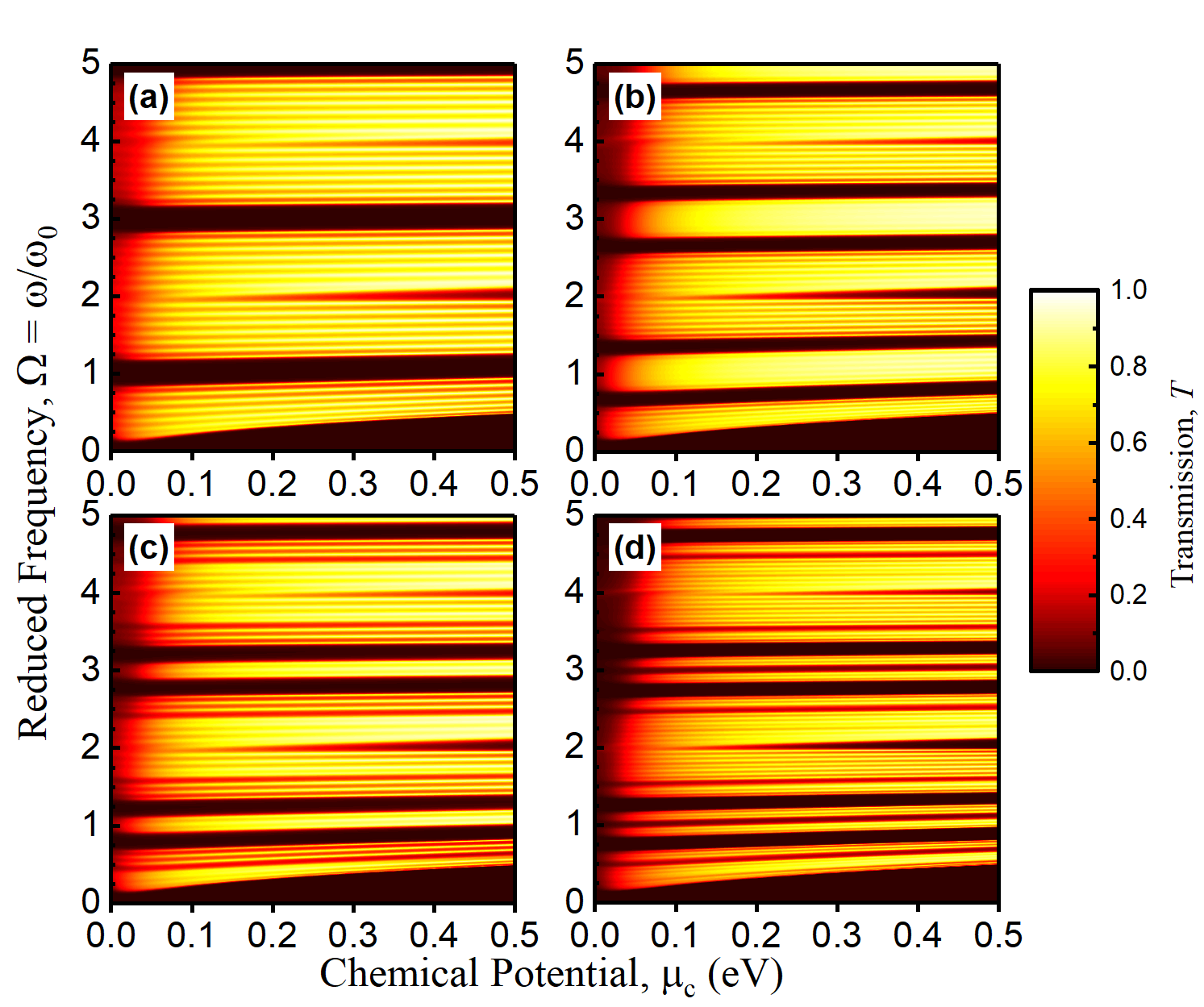}
  \caption{Transmission spectra as a function of the reduced frequency $\Omega$ and chemical potential $\mu_{c}$, normal incidence ($\theta=0^{\circ}$), for: (a) second, (b) third (both with $N=4$) and (c) fourth, (d) fifth (both with $N=3$) generations of the Fibonacci sequence.}\label{fig:OMEGAxMUxT-2345GER}
\end{figure}

Fig.\ 12 shows the transmission spectra for the second, third, fourth and fifth generations of the Fibonacci sequence. For this case we can observe the presence of both GIPBGs (in the low frequency region) and Bragg photonic band gaps. As before, the GIPBG width monotonically increases as the chemical potential $\mu_{c}$ increases, but it is independent of the Fibonacci generation index $n$. On the other hand, the width of the Bragg photonic band gap  is nearly independent of the chemical potential $\mu_{c}$. However, they get narrower and narrower as the generation index $n$ increases.

\section{Physical origin of the GIPBG}

Before concluding, let us take a closer look at the physical origin of the GIPBGs. From a mathematical point of view, the GIPBGs are associated to the trace of the transfer matrix of the system, which has two contributions: one from Bragg scatterings (depending only on the optical and geometrical parameters of the PQC) and other from the graphene surface conductance (disappearing when $\sigma_{g}$ is small)\cite{Soukoulis_PRB_2013,Madani_PhysicaB_2013}. In the low frequency region the contribution from $\sigma_{g}$, to the trace of the transfer matrix, is dominant and $|(1/2)\textrm{Tr}\left[M_n\right]|>1$ (see Figs.\ 4 and 5). As a consequence, the GIPBGs emerge. From a physical point of view, the GIPBGs are consequence of full reflection of the eletromagnetic waves. The permittivity of the building blocks $A$ and $B$ may be written as\cite{Markos_Book_2008},

\begin{equation}\label{eq:permissividade1}
\displaystyle \epsilon_{A(B)}=\epsilon_{TiO_2(SiO_2)}+i\frac{4\pi\sigma_g}{\omega}
\end{equation}

or

\begin{equation}\label{eq:permissividade2}
\displaystyle \epsilon_{A(B)}=\epsilon_{TiO_2(SiO_2)}+i\frac{4\pi}{\omega}\left[ \sigma_{g,R}+i\sigma_{g,I} \right],
\end{equation}

\noindent where $\sigma_{g,R}$ and $\sigma_{g,I}$ are the real and imaginary parts of the graphene surface conductance $\sigma_g$, respectively. In the low frequency region the  imaginary part of $\sigma_g$ dominates, as shown in Fig.\ 2, and the permittivity takes the form,

\begin{equation}\label{eq:permissividade3}
\displaystyle \epsilon_{A(B)}\approx\epsilon_{TiO_2(SiO_2)}-\frac{4\pi\sigma_{g,I}}{\omega}.
\end{equation}

\begin{figure}[h]
  \centering
  \includegraphics[width=.5\textwidth]{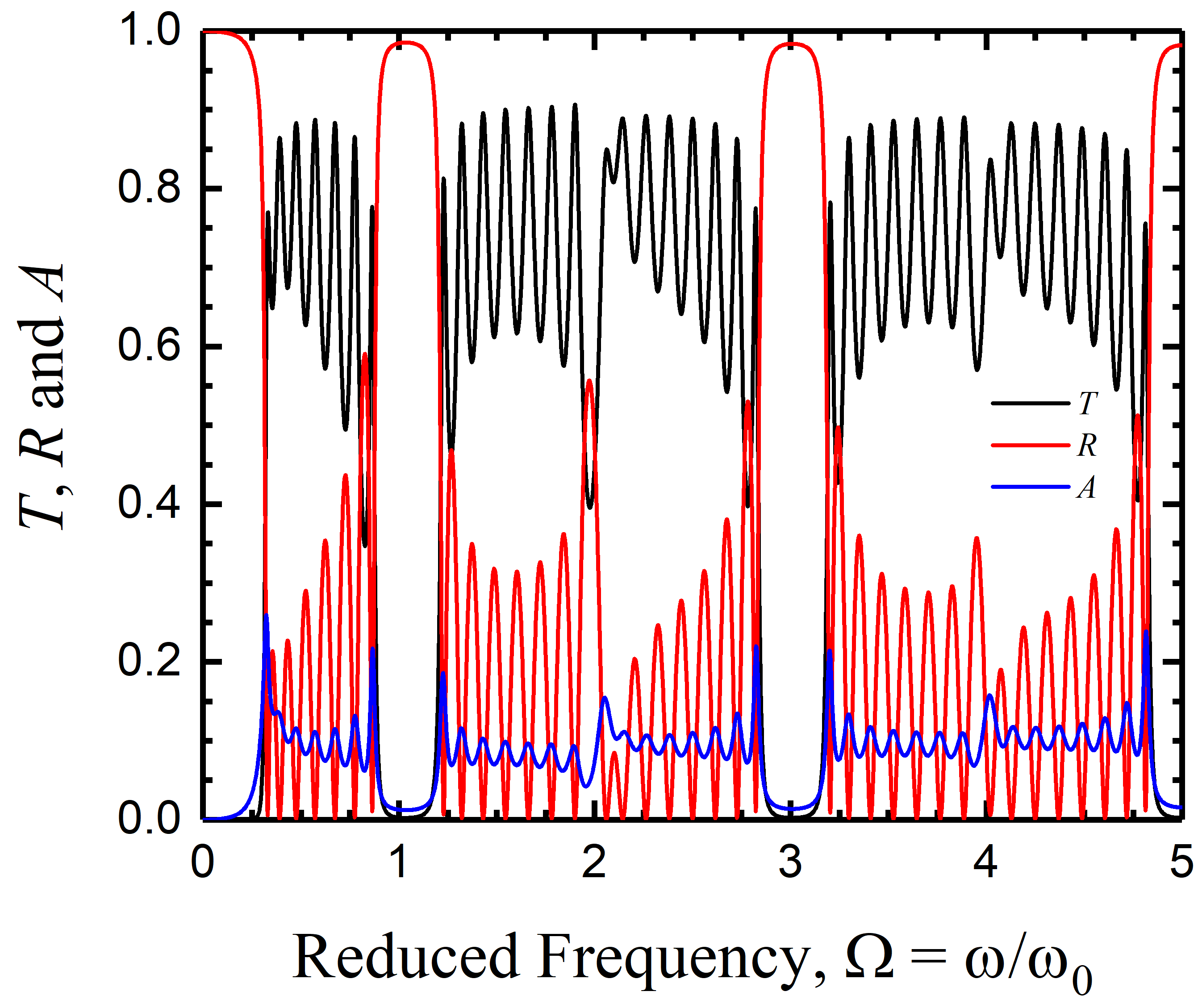}
  \caption {Transmission (black line), reflection (red line) and absorption (blue line) spectra as a function of the reduced frequency $\Omega=\omega/\omega_{0}$, with normal incidence, for the periodic case. The physical parameters used here are the same of Fig.\ 3.}
\end{figure}

\noindent Therefore, once the imaginary part of $\sigma_g$ dominates, full reflection takes place in the low frequency region and no eletromagnetic wave is transmitted. In Fig.\ 13 are shown the coefficients of transmission $T$, reflection $R$ and absorption $A$ of the incident electromagnetic wave for the periodic case. From Fig.\ 13 we can conclude that the physical origin of the GIPBGs is mainly due to full reflection, with absorption playing a small role. Regarding the applications, PBGs present a number of options to tailor the flow of light. For instance, photonic waveguides, photonic fibers, photonic microcavities, among others (see for example Ref.\ [23] and references therein).

\section{Conclusions}\label{sec:Conclusions}
In summary, we have employed a transfer matrix treatment to study the propagation of light waves in quasiperiodic dieletric multilayers with graphene embedded. We calculated their dispersion and transmission spectra to investigate the effects of the graphene monolayers and quasiperiodic disorder on the system physical behavior. The quasiperiodic disorder considered is due to the Fibonacci sequence which was used to build the dielectric multilayer composed by two building blocks, with graphene embedded, namely, silicon dioxide (building block $A=\textrm{SiO}_{2}$) and titanium dioxide (building block $B=\textrm{TiO}_{2}$). In our numerical results we show that the presence of the graphene monolayers between the dielectric layers reduces the transmissivity on the whole range of frequency and induces a transmission gap in the low frequency region. These \textit{graphene induced photonic band gaps} (GIPBG) are unusual because they do not come from Bragg reflections. As a matter of the fact, the physical origin of the GIPBGs is mainly associated with full reflection of the eletromagnetic waves, with absorption playing a small role, in the low frequency region as illustrated in Fig.\ 13. Regarding the polarization of the light wave, our numerical results show that the transmission coefficient is higher for the TM case than for the TE one. We also conclude from our numerical results that the GIPBGs do not depend on the polarization of light nor on the Fibonacci generation index $n$. In fact, the GIPBGs are \textit{omnidirectional photonic band gaps}, i.e., the light cannot propagate in this structures for $\Omega\leq 0.38$ whatever is the angle of incidence. On the other hand, Bragg photonic band gaps depend on the polarization of light and Fibonacci generation index $n$. In particular, there is no Bragg photonic band gap for the TM case. As the generation index increases $n$, the Bragg photonic band gaps get narrower and narrower. Also, the plot of the transmission spectra versus the chemical potential shows that there is only one photonic band gap in the low-frequency region (GIPBG). Furthermore, it shows that we can adjust the width of the photonic band gap in a quasiperiodic photonic crystal by tuning of the chemical potential $\mu_{c}$ with a gate voltage. In fact, the GIPBGs width monotonically increases as the chemical potential $\mu_{c}$ increases. All physical phenomena presented here can be experimentally tested and we hope that our numerical results will stimulate experimental groups to pursue them.

\begin{acknowledgments}
\noindent We would like to thank the Brazilian Research Agencies CNPq, CAPES and INCT of Space Studies for their financial support.
CGB and LFCP acknowledge financial support from the Brazilian government agency CAPES for the project ``Physical properties of nanostructured materials'' (Grant 3195/ 2014) via its Science Without Borders program. We also would like to thank the two anonymous referees for their valuable comments which helped to improve the manuscript.
\end{acknowledgments}


\newpage

\begin{thebibliography}{00}

\bibitem{Yablonovitch_PRL_1987} J. E. Yablonovitch, Phys. Rev. Lett. \textbf{58}, 2059 (1987).

\bibitem{John_PRL_1987} S. John, Phys. Rev. Lett. \textbf{58}, 2486 (1987).

\bibitem{Petritsch_Book_2017} K. Petritsch, \textit{Photonic Crystals - Introduction, Theory and Applications} (Arcler Press LLC, 2017).

\bibitem{Gong_Book_2014} Q. Gong, and X. Hu (Eds.), \textit{Photonic Crystals: Principles and Applications} (Pan Stanford, 2014).

\bibitem{Laine_Book_2010} V. E. Laine (Ed.), \textit{Photonic Crystals: Fabrication, Band Structure and
Applications} (Nova Science Pub. Inc., 2010).

\bibitem{Sukhoivanov_Book_2009} I. A. Sukhoivanov, and I. V. Guryev, \textit{Photonic Crystals: Physics and Practical Modeling} (Springer, 2009).

\bibitem{Joannopoulos_Book_2008} J. D. Joannopoulos, S. G. Johnson, J. N. Winn, and R. D. Meade, \textit{Photonic Crystals: Molding the Flow of Light} (Princeton University Press, 2nd ed., 2008).

\bibitem{Sakoda_Book_2004} K. Sakoda, \textit{Optical Properties of Photonic Crystals} (Springer, 2nd ed., 2004).

\bibitem{Markos_Book_2008} P. Marko\u{s}, and C. M. Soukoulis, \textit{ Wave Propagation: From Electrons to Photonic Crystals and Left-Handed Materials} (Princeton University Press, 2008).

\bibitem{Novoselov_Science_2004} K. S. Novoselov, A. K. Geim, S. V. Morozov, D. Jiang, Y. Zhang, S. V. Dubonos, I. V. Grigorieva, A. A. Firsov, Science	\textbf{306}, 666 (2004).

\bibitem{Wolf_Book_2016} E. L. Wolf, \textit{Graphene: A New Paradigm in Condensed Matter and Device Physics} (Oxford University Press, 2016).

\bibitem{Torres_Book_2014} L. E. F. Torres, S. Roche, and J.-C. Charlier, \textit{Introduction to Graphene-Based Nanomaterials: From Electronic Structure to Quantum Transport} (Cambridge University Press, 2014).

\bibitem{Wolf_Book_2014} E. L. Wolf, \textit{Applications of Graphene: An Overview}  (SpringerBriefs in Materials Series, Springer, 2014).

\bibitem{Aoki_Book_2013} H. Aoki, M. S. Dresselhaus (Eds.), \textit{Physics of Graphene} (NanoScience and Technology, Springer, 2013).

\bibitem{Rao_Book_2013} C. N. Rao, and A. K. Sood (Eds.), \textit{Graphene: Synthesis, Properties, and Phenomena} (Wiley-VCH, 2013).

\bibitem{Pereira2013} L. F. C. Pereira, and D. Donadio, Phys. Rev. B.  \textbf{87}, 125424 (2013).

\bibitem{Xu2014} X. Xu, L. F. C. Pereira, Y. Wang, J. Wu, K. Zhang, X. Zhao, S. Bae, C. T. Bui, R. Xie, J. T. L. Thong, B. H. Hong, K. P. Loh, D. Donadio, B. Li, and B. \"Ozyilmaz, Nat. Commun. \textbf{5}, 3689 (2014). 

\bibitem{Sharon_Book_2015} M. Sharon, M. Sharon, A. Tiwari , and H. Shinohara (Eds.), \textit{Graphene: An Introduction to the Fundamentals and Industrial Applications} (Advanced Material Series, Wiley-Scrivener, 2015).

\bibitem{Souza_Book_2014} F. D'Souza, and K. M. Kadish (Eds.), \textit{Handbook of Carbon Nano Materials - Vol. 5: Graphene - Fundamental Properties} and \textit{Vol. 6: Graphene - Energy and Sensor Applications} (World Scientific Series on Carbon Nanoscience, World Scientific Publishing Company, 2014).

\bibitem{Warner_Book_2012} J. H. Warner, F. Schaffel, M. Rummeli, A. Bachmatiuk, \textit{Graphene: Fundamentals and emergent applications} (Elsevier, 2012).

\bibitem{Nikitin_JO_2013} A. Y. Nikitin, S. A. Maier, and L. M.-Moreno, J. Opt. \textbf{15}, 110201 (2013); and references therein.

\bibitem{Abajo_Science_2013} F. J. G. Abajo, Science \textbf{339}, 917 (2013). 

\bibitem{Soukoulis_PRB_2013} Y. Fan, Z. Wei, H. Li, H. Chen, and C. M. Soukoulis, Phys. Rev. B \textbf{88}, 241403(R) (2013).

\bibitem{Shechtman_PRL_1984} D. Shechtman, I. Blech, D. Gratias, and J. W. Cahn, Phys. Rev. Lett. \textbf{53}, 1951 (1984).

\bibitem{Limonov_Book_2016} M. F. Limonov, R. M. Rue (Eds.), \textit{Optical Properties of Photonic Structures: Interplay of Order and Disorder} (Series in Optics and Optoelectronics, CRC Press, 2016). 

\bibitem{Fujiwara_Book_2014} T. Fujiwara, Y. Ishii (Eds.), \textit{Quasicrystals} (Elsevier Science, 2014).

\bibitem{Negro_Book_2013} L. D. Negro (Ed.), \textit{Optics of Aperiodic Structures: Fundamentals and Device Applications} (Pan Stanford, 2013).

\bibitem{Suck_Book_2010} J.-B. Suck, M. Schreiber, and P. H\"aussler (Eds.), \textit{Quasicrystals: An Introduction to Structure, Physical Properties and Applications} (Springer Series in Materials Science, Springer, 2010). 

\bibitem{Barber_Book_2008} E. M. Barber, \textit{Aperiodic Structures in Condensed Matter: Fundamentals and Applications} (Condensed Matter Physics, CRC Press, 2008).

\bibitem{Janssen_Book_2008} T. Janssen, G. Chapuis, M. Boissieu, \textit{Aperiodic Crystals: From Modulated Phases to Quasicrystals} (Oxford University Press, 2007).

\bibitem{Shechtman_Nobel_2011} Page of 2011 Nobel Prize in Chemistry awarded to Dan Shechtman: \verb"http://www.nobelprize.org/nobel_prizes/chemistry/"\newline \verb"laureates/2011/press.html".

\bibitem{Albuquerque_Book_2004} E. L. Albuquerque, and M. G. Cottam, \textit{Polaritons in Periodic and Quasiperiodic Structures} (Elsevier, 2004).

\bibitem{Munzar} D. Munzar, L. Bocaek, J. Humlicek, and K. Ploog. J. Phys.: Condens. Matter \textbf{6}, 4107 (1994).

\bibitem{Vardeny_NP_2013} Z. V. Vardeny, A. Nahata, and A. Agrawal, Nat. Photonics \textbf{7}, 177 (2013).

\bibitem{Macia_RPP_2006} E. Maci\'a, Rep. Prog. Phys. \textbf{69}, 397 (2006).

\bibitem{Madani_PhysicaB_2013} A. Madani, and S. R. Entezar, Physica B \textbf{431}, 1 (2013).

\bibitem{Ning_JO_2014} R. Ning, S. Liu, H. Zhang, X. Kong, B. Bian, and J. Bao, J. Opt. \textbf{16}, 125108 (2014).

\bibitem{Farhat_OE_2013} M. Farhat, C. Rockstuhl, and Hakan Ba\u{g}c{i}, Opt. Express \textbf{21}, 12592 (2013).

\bibitem{Aliofkhazraei_Book_2016} M. Aliofkhazraei, N. Ali, W. I. Milne, C. S. Ozkan, S. Mitura, and J. L. Gervasoni (Eds.), \textit{Graphene Science Handbook: Electrical and Optical Properties} (CRC Press, 2016).

\bibitem{Costa_JPCM_2013} C. H. O. Costa, and M. S. Vasconcelos, J. Phys.: Condens. Matter \textbf{25}, 286002 (2013).

\bibitem{Zaleki_JNanophoton_2016} Z. Saleki, S. R. Entezar, and A. Madani, J. Nanophoton. \textbf{10}, 036010 (2016). 

\bibitem{Zhan_JPCM_2013} T. Zhan, X. Shi, Y. Dai, X. Liu, and J. Zi. J. Phys.: Condens. Matter \textbf{25}, 215301 (2013).

\bibitem{Brandao_OM_2015} E. R. Brand\~ao, C. H. Costa, M. S. Vasconcelos, D .H. A. L. Anselmo, and V. D. Mello, Opt. Mater. \textbf{46}, 378 (2015).

\bibitem{Vasconcelos_PRB_1999} M. S. Vasconcelos, and E. L. Albuquerque, Phys. Rev. B \textbf{49}, 11128 (1999).

\end{thebibliography}
\end{document}